\documentclass[aps,pre,reprint, amsmath, amssymb,superscriptaddress]{revtex4-1}
\usepackage{morefloats}
\usepackage{bm}
\newcommand{\beq}{\begin{equation}}
\newcommand{\eeq}{\end{equation}}

\usepackage[retainorgcmds]{IEEEtrantools}
\usepackage{graphicx,tikz,placeins}
\usepackage{mathrsfs}
\usepackage{amsmath,amssymb,amsfonts,physics}
\usepackage{color}
\usepackage{float}
\usepackage{times,txfonts}
\usepackage{nicefrac}
\usepackage{ulem}
\usepackage[colorlinks=true,linkcolor=blue,urlcolor=blue,citecolor=blue,pdfusetitle]{hyperref}
\usepackage{physics}
\usepackage{soul}

\usepackage{cancel}

\begin{document}

\title{ Thermodynamics of underdamped Brownian collisional engines: New insights
and resonant phenomena }

\author{Gustavo A. L. For\~ao}
\affiliation{Universidade de São Paulo,
Instituto de Física,
Rua do Matão, 1371, 05508-090
São Paulo, SP, Brazil}

\author{Fernando S. Filho}
\affiliation{Universidade de São Paulo,
Instituto de Física,
Rua do Matão, 1371, 05508-090
São Paulo, SP, Brazil}
\affiliation{UHasselt, Faculty of Sciences, Theory Lab, Agoralaan, 3590 Diepenbeek, Belgium}

\author{Bruno A. N. Akasaki}
\affiliation{Universidade de São Paulo,
Instituto de Física,
Rua do Matão, 1371, 05508-090
São Paulo, SP, Brazil}

\author{Carlos E. Fiore}
\affiliation{Universidade de São Paulo,
Instituto de Física,
Rua do Matão, 1371, 05508-090
São Paulo, SP, Brazil}

\date{\today}

\begin{abstract}
Collisional Brownian engines have been proposed as  alternatives for  nonequilibrium nanoscale engines. However, most studies have focused on the simpler overdamped case, leaving the role of inertia much less explored. In this work, we introduce the idea of collisional engines to underdamped Brownian particles, where each stage is sequentially subjected to a distinct driving force. A careful comparison between the performance of underdamped and overdamped Brownian work-to-work engines has been undertaken. The results show that underdamped Brownian engines generally outperform their overdamped counterparts. A key difference is the presence of a resonant regime in underdamped engines, in which both efficiency and power output are enhanced across a broad set of parameters. Our study highlights the importance of carefully selecting dynamics and driving protocols to achieve optimal engine performance.

\end{abstract}

\maketitle

The performance of  engines and  the search of  protocol optimizations   constitute fundamental issues in Thermodynamics since the seminal work by Sadi Carnot \cite{carnot1872reflexions,de2013equilibrium} in 1872. Notwithstanding, the construction of different and reliable setups aimed at converting one kind of energy into another one  has become more sophisticated, above all with the advent of the nanotechnology and new experimental procedures for investigating and creating  nanoscale engines. 
Contrasting with macroscopic engines, fluctuations in nanoscale system can become important, making necessary the use of stochastic methods in order to obtain the thermodynamics properties and its relationship with the system performance. Stochastic thermodynamics (ST) \cite{seifert2012stochastic,VANDENBROECK20156,PhysRevE.82.021120,articleb235c746}
constitute an unified tool for describing nanoscale systems operating far from equilibrium \cite{seifert2012stochastic,VANDENBROECK20156,PhysRevE.82.021120,articleb235c746} and tackling the role of fluctuation and dissipation. 

Recently, a collisional (or sequential) description has been proposed and extended for Brownian systems \cite{angel,Noa1,filho,mamede2022}. In its simplest version, a single particle interacts sequentially with a thermal bath and is subjected to a specific work source at each stage. This approach differs from the situation where the system interacts with all thermal baths simultaneously \cite{Duet,proesmans2017underdamped} and has  been studied in distinct  cases of non-equilibrium thermodynamics  \cite{rosas,parrondo2015thermodynamics}, open quantum systems \cite{frank,parrondo2015thermodynamics,strasberg,molitor} and information and computational processing \cite{parrondo2015thermodynamics,sagawa2014thermodynamic,RevModPhys.81.1,bennett1982thermodynamics}.
Under suitable conditions, the sequential interaction operates as a heat engine or work-to-work converter, generating useful power. 
Despite this,  its performance
can be small or strongly reduced depending on the way it is designed, including the  choice of ingredients such as period, temperatures, and the strength of driving worksources, revealing that the search of strategies and optimization routes as fundamental \cite{angel,Noa1,filho,harunari,rosas}. 

For describing the motion of particles in a colloidal environment subjected to random forces, a fundamental framework is  the Langevin equation or Fokker-Planck approach \cite{van1992stochastic,tome2015stochastic}. 
Its two employed variants, namely the underdamped and overdamped cases, capture distinct and essential aspects of particle dynamics and stochasticity. The underdamped variant emphasizes inertia and retains the correlation between particle and position, while the overdamped variant simplifies the description and is suitable for systems with rapid relaxation. Despite the extensive research about them \cite{tothova2022overdamped,PhysRevE.95.012120,PhysRevE.98.052107,fisher1999force,leibler1994moving,bodrova2016underdamped,naze2022optimal,de2020phase}, little is known about their thermodynamic implications and the influence of inertial considerations on system performance and dissipation \cite{Duet,proesmans2017underdamped}.

In this paper, we advance upon previous works \cite{filho,angel,Noa1,mamede2022} by introducing and extending the concept of collisional Brownian engines to underdamped systems. For equal temperatures at each stage, we obtain exact expressions for Thermodynamic quantities such as work, heat and dissipation,
solely expressed in terms of Onsager coefficients,
irrespective the driving protocol. For distinct driving worksources,  we provide a comparative study between underdamped and overdamped dynamics.
The underdamped case is significantly different  \cite{filho,angel,Noa1,mamede2022}  due to the presence of a  resonance phenomenon, resulting in a specific region on the phase space in which the engine operates at maximum power and maximum efficiency.
The present study sheds light about the importance of dynamics and driving protocols for
achieving optimal engine performances.


This paper is structured as follows: In Section \ref{two}, we introduce the model, the main the expressions for the underdamped model and the maximization routes. In Section \ref{three}, we compare the performance and dissipation of the engine ruled by each dynamic, as well as present the resonant phenomena in the Underdamped engine. Conclusions are drawn in Section \ref{four}.

\section{Thermodynamics of Underdamped Brownian Engines}
\label{two}

Our setup is composed of a Brownian particle sequentially placed in contact
with a given thermal reservoir and subjected to a total force $\tilde{f}_i(x,t)$ per mass at each stage $i$, $i\in \{1,2\}$. 
The former and latter contact has duration $\tau_1$ and $\tau-\tau_1$,
respectively,  
in which 
at each stroke $i$, the system dynamics is described by the equations,
\begin{equation}
\begin{cases}
\label{langevin1}
    \frac{d}{dt}\,v_i(t) = \,\overline{f}_i(x,t) - \gamma_i\,v_i(t) + \xi_i(t),\\
    \frac{d}{dt}\,x_i(t) = v_i(t),
\end{cases}
\end{equation}
Here, $\gamma_i$ represents the viscous coefficient per  mass. The stochastic forces follow standard white noise properties: $\langle \xi_i(t) \rangle = 0$ and $\langle \xi_i(t)\,\xi_j(t') \rangle = 2\,\gamma_i\,k_B\,T_i\,\delta_{ij}\,\delta(t-t')/m$. The probability distribution of the particle $P_i(x,v,t)$ at the stroke $i$ is described by  Fokker-Planck-Kramers (FPK) \cite{tome2015stochastic,PhysRevE.82.021120,PhysRevE.82.011144,angel},
\begin{equation}
\label{FK}
    \frac{\partial P_i}{\partial t} = -\left[v\frac{\partial P_i}{\partial x}+ \overline{f}_i(x,t)\frac{\partial P_i}{\partial v}+\frac{\partial J_i}{\partial v} \right],
\end{equation}
where    $\overline{f}_i(x,t)$ is decomposed in the following form $\overline{f}_i(x,t) = f_i^*(x) + \Tilde{f}_i(t)$, where $f_i^*(x) = -\kappa\,x_i/m$ depends on the position, and $\Tilde{f}_i(t)$ is time dependent and $J_i$ is a current of probability given by
\begin{equation}
    \label{currentsFK}
    J_i = -\gamma_i\,v\,P_i - \frac{\gamma_i\,k_B\,T_i}{m}\,\frac{\partial P_i}{\partial v}.
\end{equation}
At $t = \tau_1$, the particle switches to a second thermal bath at temperature $T_2$ and is subjected to a second force $\tilde{f}_2(x,t)$, which acts during $t \in [\tau_1,\tau)$. At $t = \tau$, a cycle is completed, and the particle returns to the first thermal bath with $T_1$ and $\tilde{f}_1(x,t)$, starting a new cycle. The collisional approach assumes the exchange of reservoirs happens instantaneously, effectively treating each switching as an adiabatic process.
Some remarks about Eq.~(\ref{FK}) are in order. First, the probability distribution has a Gaussian form regardless of the temperatures and drivings. Second, the specific case where $\Tilde{f}_1(t) = \Tilde{f}_2(t) = 0$ and the temperatures are equal, $T_1 = T_2$, corresponds to the Boltzmann-Gibbs distribution, describing the system in equilibrium thermodynamics. Third, even when $T_1 \neq T_2$ and/or $\Tilde{f}_1(t) \neq \Tilde{f}_2(t)$, the system evolves towards a nonequilibrium steady state (NESS). Lastly, to derive the model thermodynamics, we assume that both $P_i(v,x,t)$ and $J_i(v,x,t)$ vanish as $v \to \pm \infty$ and/or $x \to \pm \infty$. Therefore, the mean energy $U_i(t) = m\langle v_i^2 \rangle/2 + \kappa \langle x_i^2 \rangle/2$ has a time derivative expressed as the sum of two components,
\begin{equation}
\label{flaw}
    \frac{d}{dt}U_i(t) = -\left[\Dot{W}_i(t) +\Dot{Q}_i(t)\right],
\end{equation}
where the former and latter right terms denote the work (per time) done on the particle by the force $\Tilde{f}_i(t)$ and the heat flux $\Dot{Q}_i(t)$ exchanged with the thermal bath at the stroke $i$. Explicitly,
\begin{equation}
\label{work}
    \Dot{W}_i(t) = -m\,\langle v_i\rangle(t)\,\Tilde{f}_i(t),
\end{equation}
and
\begin{equation}
    \Dot{Q}_i(t) = \gamma_i\left(m\,\langle v_i^2 \rangle(t) - k_B\,T_i\right),
\end{equation}
\newline
respectively. From now on, we shall curb ourselves for  $\tau_1=\tau/2$, whose external drivings can be expressed in the following
form:
\begin{equation}
\Tilde{f}_i(t) =
\begin{cases}
    X_1\,g_1(t), & 0\,\leq\, t \,<\tau/2\\
    X_2\,g_2(t), & \tau/2\,\leq\, t \,<\tau,
\end{cases}
\end{equation}
where the $X_i$ denotes the strength of the thermodynamic force per mass acting on the system, while $g_i(t)$ defines the shape of the protocol.
By evaluating (\ref{flaw}) over a complete cycle and considering that the system returns to its initial state, one derives the first law of thermodynamics averaged over a period in NESS: $\overline{\dot{W}}_1 + \overline{\dot{W}}_2 +\overline{\dot{Q}}_1+\overline{\dot{Q}}_2=0$. Similarly, the second law of thermodynamics relates to the time evolution of entropy $S_i = -k_B,\langle \ln{P_i} \rangle$, which, together with Eq. (\ref{FK}), can be expressed through the difference  between the entropy production rate $\sigma_i(t)$ and the entropy flux $\Phi_i(t)$:

\begin{equation}
    \frac{d}{dt}S_i = \sigma_i(t) - \Phi_i(t),
\end{equation}
where
\begin{equation}
\label{epeq}
   \sigma_i(t) = \frac{m}{\gamma_i\,T_i}\int \frac{J_i^2}{P_i}\,dxdv\,\,\,\,\,\,\,\,\,\,\textrm{and}\,\,\,\,\,\,\,\,\,\, \Phi_i(t) = \frac{\Dot{Q}_i(t)}{T_i}.
\end{equation}

\noindent It is straightforward that $\sigma_i(t) \geq 0$, in accordance with the second law of Thermodynamics. By evaluating the  entropy $S_i$ over the complete period, one has that $\overline{\sigma} = \overline{\Phi}= -(\overline{\dot{Q}}_1/T_1 + \overline{\dot{Q}}_2/T_2)$, where  $ \overline{\sigma}=\left(\int_{0}^{\tau_1}\sigma_1(t) dt+\int_{\tau_1}^{\tau}\sigma_2(t) dt\right)/\tau$ and $ \overline{\Phi}=\left(\int_{0}^{\tau_1}\Phi_1(t) dt+\int_{\tau_1}^{\tau}\Phi_2(t) dt\right)/\tau$.
We can also  relate $\Bar{\sigma} $ with components $\overline{\dot{Q}}_i$'s
and $\overline {\dot W}_{i}$'s by means of the following expression
\begin{equation}
\label{cep}
    \Bar{\sigma} = \frac{4\,T^2}{4\,T^2 - \Delta T^2}\left[\ \frac{-(\overline {\dot W}_{1} + \overline {\dot W}_{2})}{T} + \frac{(\overline{\Dot{Q}}_1 - \overline{\Dot{Q}}_2)\,\Delta T}{2\,T^2}\right]\, ,
\end{equation}
where we introduced the variables $T=(T_1 + T_2)/2$ and $\Delta T = T_1 - T_2$ together the first law of Thermodynamics. Since one of our goals is to compare overdamped and underdamped setups, we shall focus on the simplest case, in which  temperatures are equal ($\Delta T=0$) and the system presents only two thermodynamic forces, closely related with $X_1$ and $X_2$. By adjusting 
$X_1$ and $X_2$, such system can operate as work-to-work converters,  generating useful power output during one of the two strokes. There are numerous examples of Brownian engines and biological physics operating as work-to-work converters \cite{PhysRevE.98.052107,fisher1999force,seifert2012stochastic,busiello2022hyperaccurate,liepelt2007kinesin,liepelt2009operation,hooyberghs2013efficiency}, where an amount of energy (e.g. chemical or mechanical) is converted into power output.

\noindent In this case, the Eq. (\ref{cep}) reduces to
\begin{equation}
\label{isothermal_entrop}
     \Bar{\sigma} =\frac{-(\overline {\dot W}_{1} + \overline {\dot W}_{2})}{T}.
\end{equation}


In order to exploit the driving changes
at each stroke, we express $g_i(t)$ in terms of its Fourier components. More specifically,
\begin{equation}
     g_1(t) = \frac{a_0}{2} + \sum_{n=1}^{\infty}a_n\cos{\left(\frac{2 \pi\,n\,t}{\tau}\right)} + b_n\sin{\left(\frac{2 \pi\,n\,t}{\tau}\right)},
     \label{d1}
\end{equation}
and
\begin{equation}
      g_2(t) = \frac{c_0}{2} + \sum_{n=1}^{\infty}c_n\cos{\left(\frac{2 \pi\,n\,t}{\tau}\right)} + d_n\sin{\left(\frac{2 \pi\,n\,t}{\tau}\right)},
      \label{d2}
\end{equation}
respectively, where the driving to be considered is characterized by obtaining the coefficients \( a_n, b_n, c_n \), and \( d_n \) under the conditions \( c_n = d_n = 0 \) for \( i = 1 \) and \( a_n = b_n = 0 \) for \( i = 2 \). It is worth mentioning that the Fourier representation ensures the boundary conditions with respect to  the probability continuity at \( t = \tau/2 \) taking into account  the system returns to the initial state at \( t = \tau \).
 By averaging Eqs.~(\ref{langevin1}), one finds the following general expression for
 $\langle v \rangle (t)$
\begin{align}
\label{meanvelocity}
    \langle v \rangle (t) =\sum_{k = 1}^{\infty}(X_1\,\cdot\,a_{1vk} &+ X_2\,\cdot\,a_{2vk} )\, \cos{\left(\frac{2 \pi\,k\,t}{\tau}\right)} \nonumber\\&+ (X_1\,\cdot\,b_{1vk} + X_2\,\cdot\,b_{2vk} )\,\sin{\left(\frac{2 \pi\,k\,t}{\tau}\right)},
\end{align}
\noindent where $a_{ivk}$ and $b_{ivk}$ correspond
to Fourier coefficients obtained for the mean velocity and depend on the driving form from Eqs.~(\ref{d1})-(\ref{d2}). Its explicit form is
shown in the Appendix \ref{appendix onsager}.  By plugging Eq. (\ref{meanvelocity})
into Eq. (\ref{work}) and averaging over the duration of each stroke, one has $\overline {\dot W}_{1}$ and $\overline {\dot W}_{2}$ given by
\begin{equation}
\label{w11}
\overline {\dot W}_{1} = - TJ_1f_1 = -T(L_{11}\,f_1^2 + L_{12}\,f_1\,f_2),
\end{equation}
\begin{equation}
\label{power}
    \overline {\dot W}_{2} = - TJ_2f_2 = -T(L_{22}\,f_2^2 + L_{21}\,f_2\,f_1),
\end{equation}
expressed in terms of thermodynamic forces $f_i = X_i/T$  ($i\in\{1,2\}$)    and $L_{ij}$'s denote Onsager coefficients, whose main expressions are listed in  Appendix  
 \ref{appendix onsager}. Note that Eq.~(\ref{cep}) can be re-expressed in terms
 of Onsager coefficients (in the regime of small $\Delta T$) as
 $\overline{\sigma} \approx J_1 f_1+J_2 f_2 + J_T f_{T}$, with  $J_T = L_{TT}f_T$ and $f_T=\Delta T/T^2$ and $L_{TT}$ is the  associated  Onsager coefficient.  Since we are focused on the isothermal case, $f_T=0$. Besides,
 in similarity with the overdamped case, there is no coupling between work fluxes and heat flux, 
 implying that $L_{1T}=L_{T1}=L_{2T}=L_{T2}=0$, consistent with the fact that such class of engines solely operates as  work-to-work converters, rather than converting heat into work output.


In order the system can operate as a work-to-work converter, it is required an amount
 of $\overline {\dot W}_{\mathrm in}=\overline {\dot W}_{i}$ be partially converted into work output  $\mathcal{P} = \overline {\dot W}_{\mathrm out}=\overline {\dot W}_{j}$, where $\mathcal{P} \geq 0 > \overline {\dot W}_{\mathrm in}$. The efficiency of conversion is thus given by
\begin{equation}
  \eta \equiv -\frac{\mathcal{P}}{\overline {\dot W}_{\mathrm in}},
\end{equation}
where $0\le \eta\le 1$. Notice that, for certain ranges of
parameters, $\overline {\dot W}_{j}$ can be greater than $\overline {\dot W}_{i}$, only meaning that in such case the amount of work $\overline {\dot W}_{\mathrm in}=\overline {\dot W}_{j}$ is
partially converted into  $\overline {\dot W}_{\mathrm out}=\overline {\dot W}_{i}$ and hence ($\Tilde{f}_i \leftrightarrow \Tilde{f}_j \Rightarrow \overline {\dot W}_{\mathrm in} \leftrightarrow \overline {\dot W}_{\mathrm out}$),  solely implying that $\eta \leftrightarrow 1/\eta$. This change of  regime is better exploited in Appendix \ref{regimeSec}. When expressed in terms of Onsager coefficients and the thermodynamic forces, $\eta$ is given by
\begin{equation}
\label{effonsager}
    \eta = -\frac{J_j f_j}{J_i f_i}=  -\frac{L_{jj}\,f_j^2 + L_{ji}\,f_i\,f_j}{L_{ii}\,f_i^2 + L_{ij}\,f_j\,f_i}.
\end{equation}

\noindent 

\subsection{Overview about system  maximizations}
There are distinct routes for the optimization of  power and efficiency in Brownian work-to-work converters. As stated in previous works \cite{filho,Noa1,Duet}, it is possible to express their optimized values in terms of Onsager coefficients. In this section,
we briefly review them. Starting with the power and considering that $f_i$ is held fixed, the engine regime is constrained between $0<f_j<|f_m|$, where $f_m=-L_{ji} f_i/L_{jj}$. The force $f^{\textrm{MP}}_j$ ensuring maximum power $\mathcal{P}_{MP}$ is given by $f_j^{\textrm{MP}} = -L_{ji} f_i/2L_{jj}$. By inserting this relation into Eqs.~(\ref{power}) and (\ref{effonsager}), one obtains  the following
expressions for quantities at maximum power:
\begin{equation}
    \mathcal{P}_{\textrm{MP}} = T\frac{\,L_{ji}^2}{4\,L_{jj}}f_i^2, \qquad
    \eta_{\textrm{MP}} = \frac{L_{ji}^2}{4\,L_{jj}\,L_{ii} - 2\,L_{ji}\,L_{ij}}.
    \label{maxi_f2}
\end{equation}
Conversely, one can search for $f^{\textrm{ME}}_j$ by optimizing the efficiency with respect to $f_j$.
\begin{equation}
  f_{j}^{ME}=\frac{L_{ii} }{L_{ij} }\left(-1+ \sqrt{1-\frac{L_{ji} L_{ij}}{L_{jj} L_{ii}}}\right)f_i,
  \label{eq:x2meta}
\end{equation}
and
\begin{equation}
  \eta_{ME}=-\frac{L_{ji}}{L_{ij}}+\frac{2L_{ii}L_{jj}}{L_{ij}^2}\left(1-\sqrt{1-\frac{L_{ij} L_{ji}}{L_{ii} L_{jj}}}\right),
  \label{etame}
\end{equation}
and $\mathcal{P}_{ME}$ is obtained by inserting $ f_{j}^{ME}$ into the expression
for ${\cal P}$.
respectively.
\begin{figure*}
    \label{fig2}
    \centering
    \includegraphics[scale=0.25]{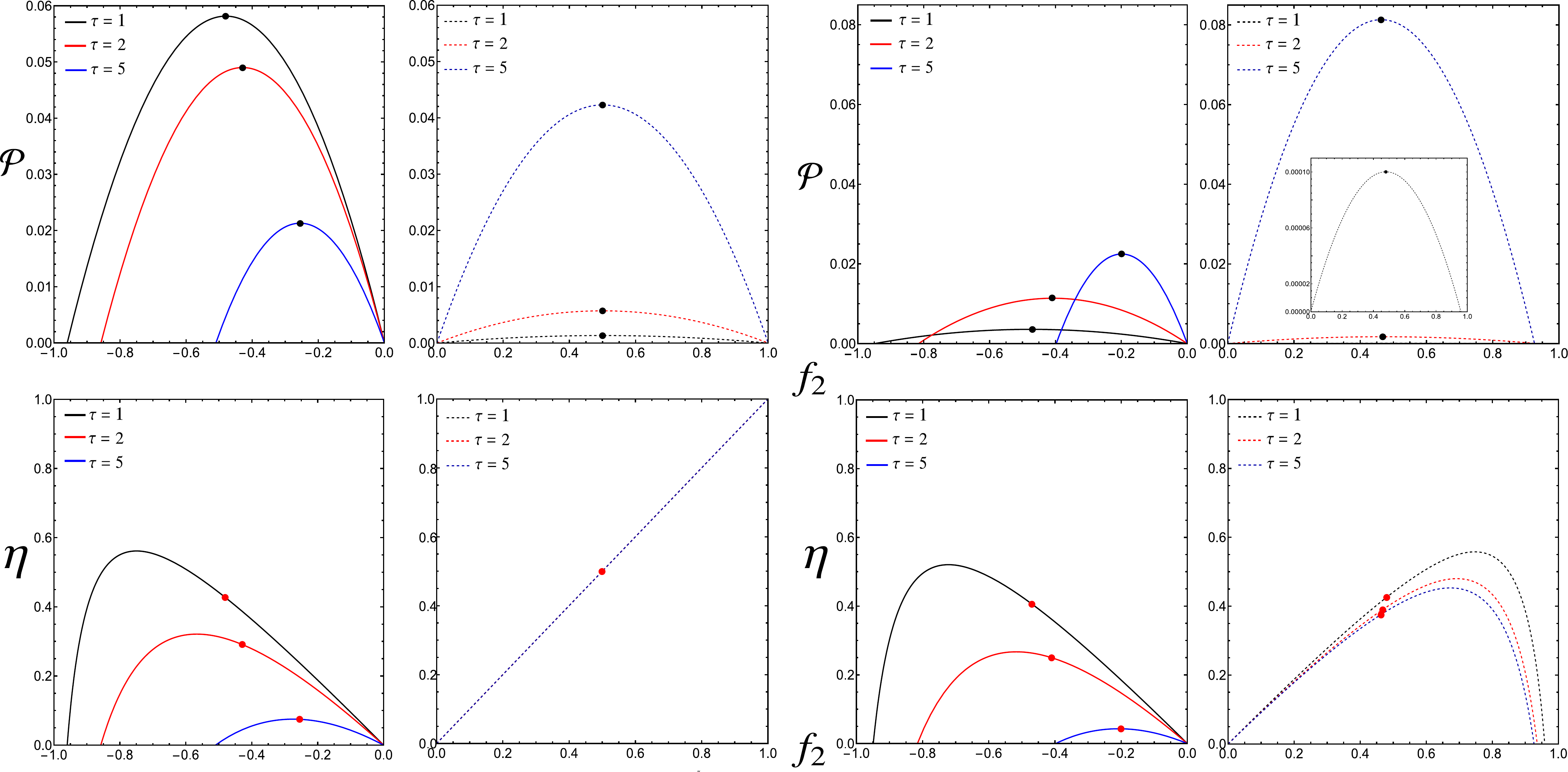}
     \caption{Depiction of the power-output ${\cal P}$ (top) and efficiency $\eta$ (bottom) as functions of $f_2$ for constant (left) and linear (right) drivings and different periods $\tau$. Dashed and continuous lines are for the underdamped and  overdamped regimes, respectively. Black and red dots represent $\mathcal{P}_{\textrm{MP}}$ and $\eta_{\textrm{MP}}$, respectively.  Parameters: $f_1 =T =1$, $\gamma = 1$ and $\kappa = 1$.}
    \label{fig1}
\end{figure*}

\section{Results}
\label{three}
Unless it has been explicitly stated, we shall adopt $k_B=m=1$, $\gamma_1 = \gamma_2 = \gamma$.
Along this section, we draw a comparison between underdamped and overdamped Brownian engines for distinct driving forces. Our analysis assumes a constant (square wave) driving, given by $g_1(t) = g_2(t)=1$, and a linear (sawtooth) driving, given by
\begin{equation}
    g_i(t) = 
    \begin{cases}
        \lambda t , & 0\,\leq\, \lambda t \,\leq\,\tau/2\\
      \lambda( t-\tau/2) \, & \tau/2\,\leq\, t \,\leq\,\tau,
    \end{cases}
\end{equation}
where $\lambda$ is a constant  introduced in such a way $g_i(t)$ is dimensionaless.
In all cases we set $\lambda=1$.
We also investigate   sinusoidal driving at each stage, as can be seen in the Appendix \ref{appendixper}. Fourier components of each driving protocol can be found in Appendix \ref{appendix onsager}.


\subsection{Dynamics Comparison: Performance Contrast}
In order to obtain a first insight about the similarities and differences
between underdamped and overdamped collisional Brownian engines, Fig.~\ref{fig1} depicts, for distinct periods
$\tau$'s, the power and efficiency versus the force  $X_2=Tf_2$ for constant and linear drivings. For simplifying matters, such first
sort of analysis will be carried out for $\gamma=\kappa=1$.
In both cases, the system can operate as a work-to-work converter by choosing 
$f_2$ constrained between
$0$ and $|f_m|$. We point out some remarkable differences between them, which  can be understood by resorting to Onsager coefficients.
For the (underdamped) constant protocol, expressions for Onsager coefficients listed in Appendix  \ref{appendix onsager} show  that $L_{11} = L_{22}=-L_{12}=-L_{21}$. As a consequence,  the engine regime is delimited
by $0\le f_2\le f_m=f_1$
(for all values of $\tau$), 
and both ${\cal P}$ and $\eta$ acquire simple forms, given by ${\cal P}=L_{22}f_2(f_2-f_1)$ and $\eta=f_2/f_1$, respectively, such latter
solely depending on the ratio between forces,
followed by maximum efficiency and power points yielding at $f^{\rm ME}_2=f_1$
and   $f^{\rm MP}_2=f_1/2$, respectively, irrespective the values
of $\kappa,\omega$ and $\tau$.
Maximum values read $\eta_{MP}=1/2,\eta_{ME}=1,{\cal P}_{MP}=TL_{22}f_1^2/4$ promptly obtained from Eqs.~(\ref{maxi_f2})-(\ref{etame}). While the former two are independent
on the period, there is  an optimal $\tau$ providing maximum ${\cal P}_{MP}$ because  $L_{ij}$'s exhibit a non monotonous behavior as $\tau$ is raised.
In particular, for the sort of parameters
it is peaked at $\tau_o=6.311...$, whose maximum  ${\cal P}_{MP}=0.05148$.
All such features are quite different from the overdamped
case whose engine regime is strongly dependent on $\tau$ in which both power and efficiencies decrease as $\tau$ is raised (see e.g. left panels).  Conversely, ${\cal P}_{MP}$ decreases
as $\tau$ is raised for the overdamped case.


The linear case is more revealing. Although Onsager coefficients also satisfy the relations $L_{11} = L_{22}$ and $L_{12}=L_{21}$, one has  $L_{11}\neq -L_{12}$, and  $L_{11}/L_{22}$  increases ``faster" than $L_{12}/L_{21}$ 
as $\tau$ is raised. As a first consequence,  $f_m$ behaves differently from the constant case and  smoothly decreases as $\tau$ is raised. Also,  ${\cal P}_{MP},\eta_{MP},{\cal P}_{ME}$ and $\eta_{ME}$ are also  differently from the constant case and depend on the interplay between $\kappa$
and $\tau$. For $\kappa=f_1=1$, ${\cal P}_{MP}$ is maximum for $\tau_o=7.345...$
whose associate power and efficiency are given by ${\cal P}_{MP}=0.170...$
and 
$\eta_{MP}=0.2765...$, respectively. The existence of an optimal $\tau_o$
ensuring maximum ${\cal P}_{MP}$ also yields in the overdamped case, yielding at $\tau_o=4.695...$ with a lower   maximum power  ${\cal P}_{MP}=0.0224...$ (evaluated for $f_1=\gamma=1$).

Above findings can be understood in terms of some heuristic (no rigorous) arguments. We first note that, for a given protocol, strength
driving forces in overdamped and underdamped cases operate at  opposite and same directions in the work-to-work regime,
respectively. This is due  there is no need for an extra force to bring the system back to its initial position in the latter case, while in the former (overdamped) case, due  to the absence of a restoring potential, it is essential for the system  have two drivings acting in different directions in order to generate useful power.
Such findings seem to be  general and also verified for periodic protocols, but there are some subtleties for sinusoidal drivings, in part  due to the phase difference between drivings (see Appendix \ref{appendixper}). 
Another difference between underdamped and overdamped relies the interplay between $f_m$ and $\tau$. Although it can be  understood directly from Onsager coefficients, our heuristic argument states that  the mean velocity  aligns to the external force in both stages  (for the overdamped case) for  sufficiently long periods and hence no useful power output is generated. This is quite different for
the underdamped case, due to the influence of restoring force
and its interplay with period and driving.

\subsection{Deterministic Resonant Phenomena in Underdamped Dynamics}
\label{resonance}

The presence of a restoring potential in the underdamped case results in a resonant phenomena when an external force drives on the system and can significantly influence the system performance
and the behavior of  thermodynamic quantities. 
Such phenomena shares some similarities with the so called Stochastic Resonance (SR),  verified recently in an experimental probe containing interacting resonant Brownian \cite{luo2022perfect}. In the present case, our goal is to study the effects of a deterministic resonance due to an external driving at each stroke. 

\begin{figure}[t!]
    \centering
    \includegraphics[scale = 0.25]{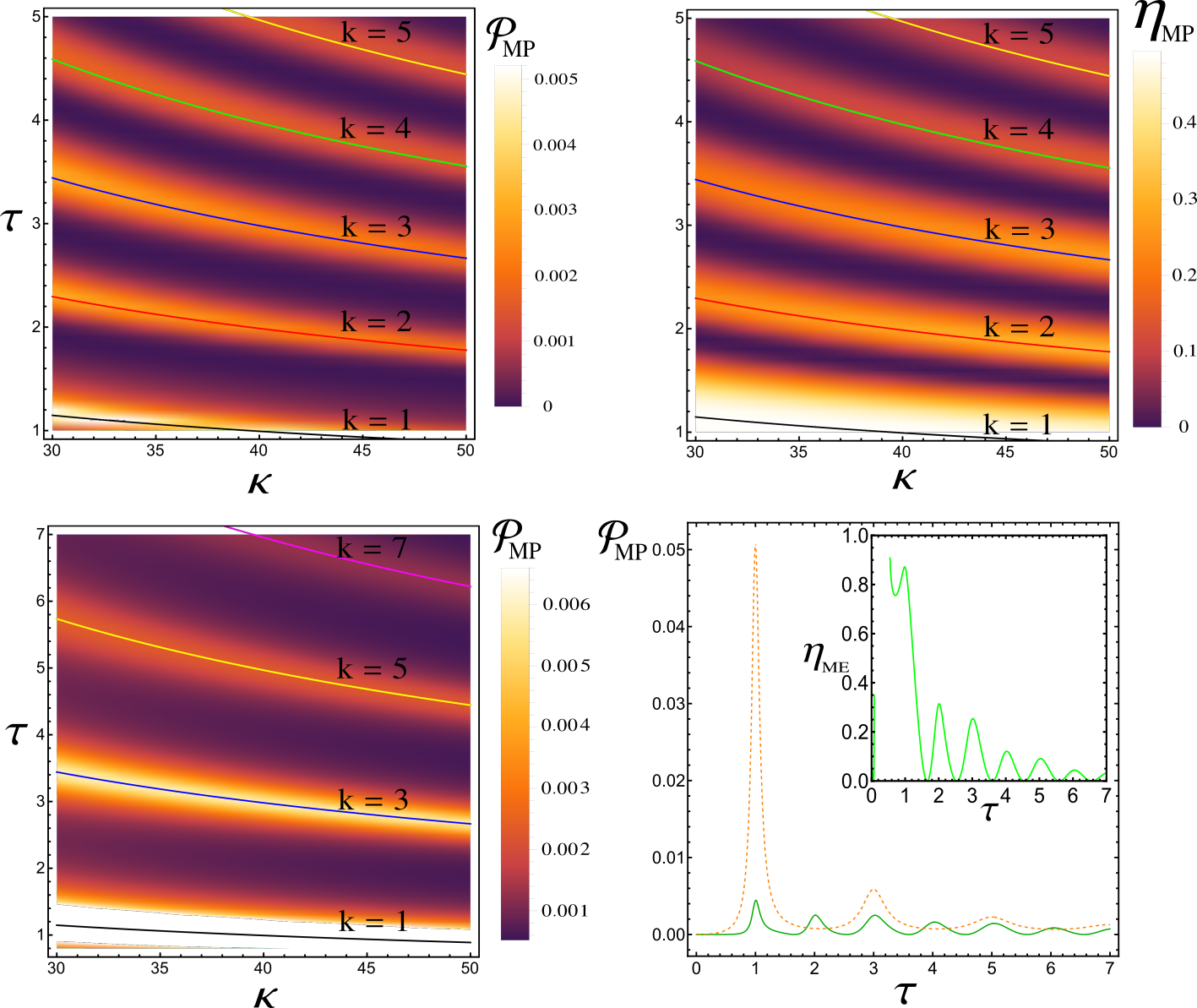}
    \caption{Top panels: Depiction of the maximum power ${\cal P}_{MP}$ (left) and efficiency at maximum power $\eta_{MP}$ (right) heat maps for linear drivings. Left bottom
    panel show the same, but for constant drivings. Continuous lines show the
    resonance lines according to Eq.~(\ref{reson}). All maximizations have been carried out with respect to $f_2$. Since $\eta_{MP}=1/2$ for constant drivings, heat maps in such case has not been shown.
    Right bottom panels: Plot of ${\cal P}_{MP}$ for constant (dashed) and linear (continuous) drivings
    for $\kappa=4\pi^2$. Inset: the same but
    for $\eta_{ME}$ (linear).
   Parameters: $f_1 =T= 1$ and $\gamma = 1$. }
    \label{hm}
\end{figure}

We start this section by deriving the resonant  $\kappa_{\textrm{res}}$  in terms of the system's parameters. The external driving is also a periodic force with  total period $\tau$ and frequency $\omega = 2\pi/\tau$. The $k$-th contribution to the Fourier series of the mean velocity (\ref{meanvelocity}) is expressed as:
\begin{align}
 \langle v(t)\rangle^{k-\textrm{th}}& = (X_1\cdot a_{1vk} \,+\, X_2\cdot a_{2vk} )\, \cos{\left(\frac{2 \pi\,k\,t}{\tau}\right)}\nonumber\\
& +\,(X_1\cdot b_{1vk}\,+\, X_2 \cdot b_{2vk} )\,\sin{\left(\frac{2 \pi\,k\,t}{\tau}\right)}.
\label{velocity_fourier}
 \end{align}
The resonant phenomenon is characterized by a maximum of   amplitude of 
$\langle v(t)\rangle^{k-\textrm{th}}$ with respect to the time $t$ and $\kappa$, described by the following relation. 
\begin{equation}
   \kappa_{\textrm{res}} = \left( \frac{2\,\pi\,k}{\tau}\right)^2, \qquad k \quad {\rm integer}
   \label{reson}
\end{equation}
Such result is protocol-independent,  solely depending on the natural frequency $\omega_0=2\pi/\tau$, and hence valid to a generic  external driving.  The relationship between resonance and the system performance
is depicted in
Fig.~\ref{hm}  for constant and linear drivings. 
We first note that it can remarkably influence the system performance, not only as forces $f_1/f_2$ are varied (not shown), but also for their optimized values ${\cal P}_{MP}/\eta_{MP}$ and ${\cal P}_{ME}/\eta_{ME}$ evaluated at $f_2=f^{\rm MP}_{2}$ and $f_2=f^{\rm ME}_{2}$, respectively.  
In both cases, resonances lines (continuous lines in heat maps) obey  Eq.~(\ref{reson})   and are followed by the increase of power and efficiency
for constant and linear drivings for odd 
and any integer  $k$, respectively. A direct inspection of Eq.~(\ref{reson}) for constant drivings  reveals that
$\langle v(t)\rangle^{k-\textrm{th}}<<1$ for even $k$, leading to strongly small performances in such a case. Bottom left panels
reinforce such findings for $\kappa=4\pi^2$ in which $\tau_{res}=k$. Also, in accordance with that stated in the previous section, $\eta_{MP}$ and $\eta_{ME}$ are always
constant in such case for constant protocols. 
As a complementar analysis, we also compare numerically the performances at the resonance and out of the resonance. For example,  for  $\tau=5$, $\kappa_{\rm res}=4\pi^2/25$  and $\kappa=1$, both constant (linear) drivings provide
the system operates with larger performances in the resonance ${\cal P}_{MP}=5,12.10^{-2}(9,91.10^{-2})$ than out the resonance ${\cal P}_{MP}=4,23.10^{-2}(8,12.10^{-2})$ , respectively, but in both cases, efficiencies are close to each other  $\eta_{MP}=1/2(0.38)$. 
Hence, the existence of resonance in the underdamped case can confer a remarkable advantage over the overdamped dynamics, not only because it is protocol-independent but also it can enhance power  without sacrificing efficiency.
However, a common trait to all resonance patterns is that they are signed by increasing dissipation (see e.g. right panels in the Fig.~\ref{epfig}). Similar findings
are also viewed for  harmonic drivings, as depicted in Appendix \ref{appendixper} for drivings  dephased of $\pi$ at each stroke. As
a consequence, resonance lines and maximum ${\cal P}_{MP}$ and $\eta_{ME}$ are also half period translated (see e.g. Fig.~\ref{hmsin}).
As a last comment, although resonances points are very close to the peaks of ${\cal P}_{MP}$ and $\eta_{ME}$ (inset),
they do not precisely coincide (although differences typically yield
at the third decimal level).

\begin{figure}[h]
    \centering
        \includegraphics[scale=0.26]{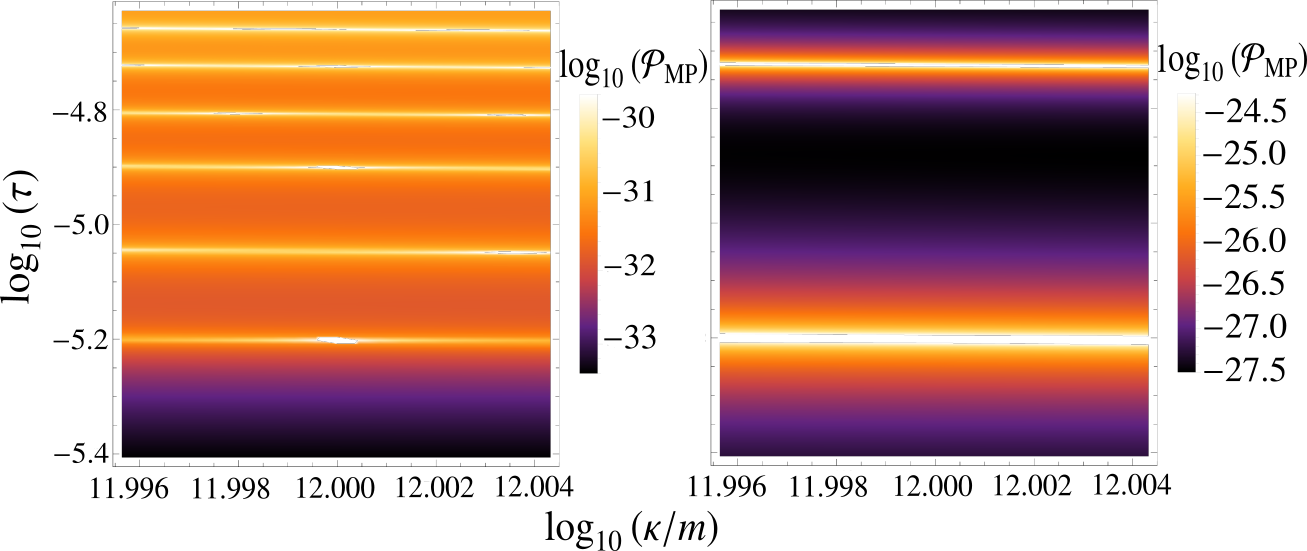}
    \caption{The power output for the linear (left) and constant (right) protocols in log base 10 scale considering some experimental quantities. Here, we use $m\approx 10^{-18}\textrm{ kg}$ and $\gamma \approx 10^{-20}\textrm{s}^{-1}$.}
    \label{resonance_real_heatmap}
\end{figure}

We close this section by  estimating how resonant phenomena can be feasible from an experimental point of view. In principle, our framework might be tested experimentally in an optical tweezers systems, in which the harmonic potential and the external drives can be generated via controlled eletric fields \cite{PhysRevA.103.013110,ghislain1994measurement,davis2007brownian,padgett2010optical}. 
For that, we take some values for laboratory quantities: $m\approx 10^{-18} \textrm{ kg}$, $\gamma \approx 10^{-20} {\rm s}^{-1}$ by imposing the thermodynamics force $X_1 \approx 0.003 \textrm{ fN/kg}$ in the underdamped case ($m/\gamma>>1$) and in room temperature with $\kappa \approx 1 \mu\textrm{N/m}$. For such values,  the resonant regime is peaked at $X_2 \approx 0.5 \textrm{fN/kg}$ in the period $\tau \approx 6 \mu \textrm{s}$ in the constant case. Fig.~\ref{resonance_real_heatmap} depicts the heat maps for constant and  linear  protocols for parameters  described above. As in Fig.~(\ref{hm}), resonance positions are coincident in both constant and linear drivings, consistent to be protocol independent. Also in accordance to
previous results, results linear drivings present more resonance lines. Although heat maps suggest that resonances can be  experimentally verified, it requires a fine tuning of model parameters, indicating that a small perturbation of parameters (e.g. the time cycle and/or in the harmonic potential)   can be drop the particle from  the resonance region. 

\begin{figure*}
    \centering
    \includegraphics[scale=0.35]{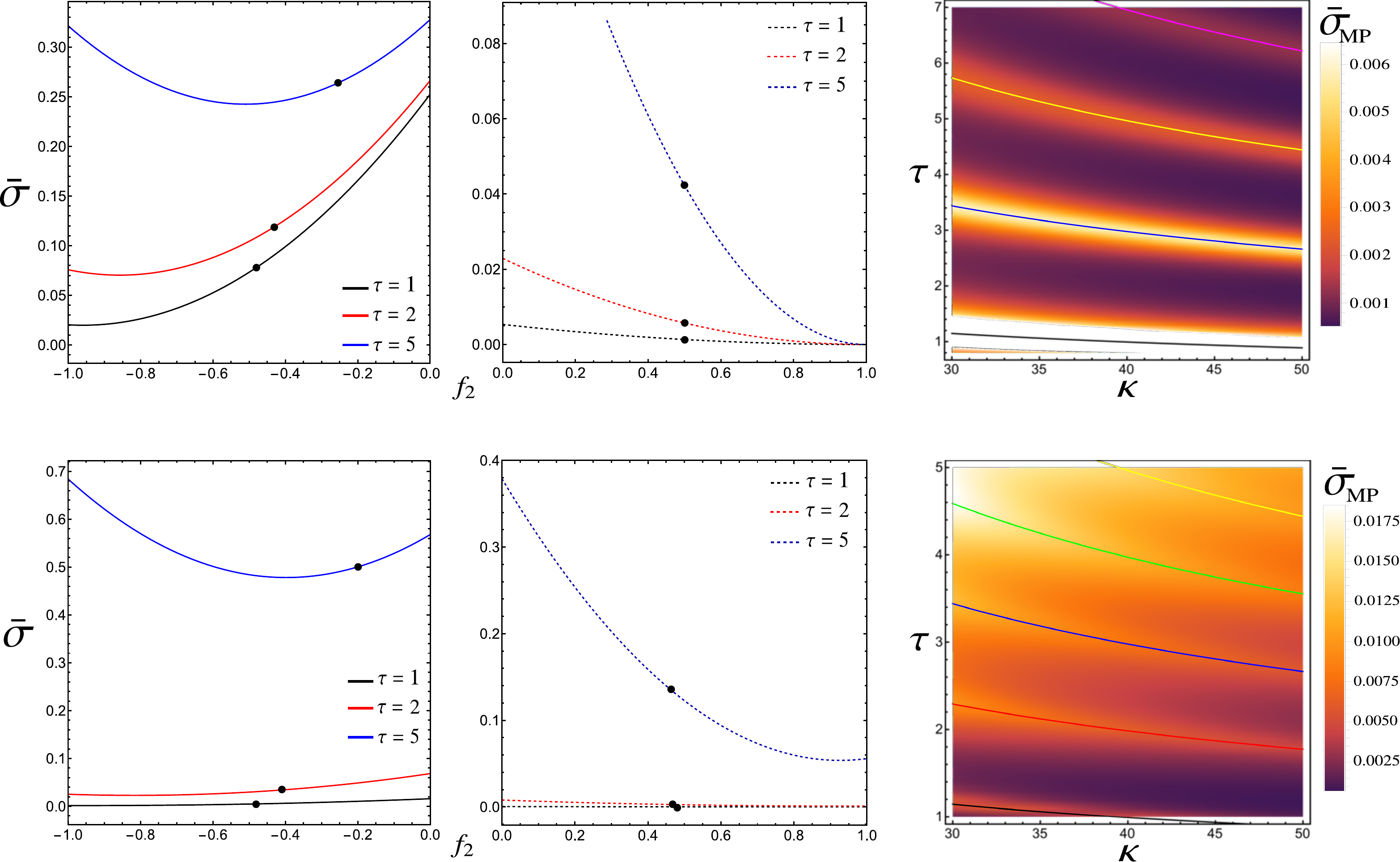}
     \caption{Left and center panels depict  the  mean entropy production $\Bar{\sigma}$ for the overdamped (left) and underdamped (center) versus $f_2$ for the same range, resonant lines and parameters as in Fig.~(\ref{performance}). Black dots represents attempt to the maximum power. Right panels show  entropy production heat maps
     for the same parameters from Fig.~(\ref{hm}).
     Top and bottom panels attempt to the constant and linear drivings, respectively.
     Parameters: $f_1 = 1$, $\gamma = 1$ and $T = 1$ in all plots and $\kappa = 1$ for the 2D plots.}
    \label{epfig}
\end{figure*}

\subsection{Entropy Production}

\subsubsection{Dynamics Comparison and dissipation at resonant regimes}

The entropy production serves as a key indicator of system dissipation, being investigated in distinct contexts, such as in the existence of different trade-offs between dissipation and fluctuation,  expressed via  TURs \cite{barato2015thermodynamic,otsubo2020estimating,hasegawa2019uncertainty,proesmans2017discrete},  its  interplay or compromise with power and efficiency in the context of thermal engines \cite{proesmans2016,Noa1,filho,angel,mamede} 
or even its usage for the characterization of phase transition regimes \cite{noa100,aguilera2023nonequilibrium,loos2023long}. Despite  it is commonly
desired to minimizing entropy production and simultaneously maximizing power and/or efficiency, the second law of thermodynamics precludes all of them be simultaneously satisfied. With this in mind, this section is aimed at depicting the entropy production behavior 
of our collisional
(underdamped Brownian) engine regimes, above all at the resonance regimes, as well as drawing a comparison with overdamped dynamics.
In terms of Onsager coefficients, Eq.~(\ref{isothermal_entrop}) 
acquires the form ${\bar \sigma}=L_{11}f_1^2+L_{22}f_2^2+(L_{12}+L_{21})f_1f_2$. 
In order to obtain a first insight, Fig.~\ref{epeq} depicts, for the same parameters as in Figs.~\ref{fig1}
and \ref{hm}, the behavior of entropy production ${\bar \sigma}$ versus $f_2$ and  ${\bar \sigma}_{MP}$ heat maps, evaluated at $f_2=f^{\rm MP}_2$ respectively.
Starting with constant  protocol, it acquires a simple form  ${\bar \sigma}=L_{22}(f_2-f_1)^2$, meaning that its minimum value ${\bar \sigma}_{mS}=0$ coincides the maximum efficiency point $f_{2}^{ME}=f_1$, irrespective the period and model parameters. This is different from the overdamped case, not only because ${\bar \sigma}_{mS}$ increases
as $\tau$ is raised (overdamped) but also 
the underdamped case is characterized by $0={\bar \sigma}_{mS}\neq {\bar \sigma}_{m}>0$  (for the overdamped case one has ${\bar \sigma}_{m}={\bar \sigma}_{mS}>0$).
Another important comparison relies the behavior ${\bar \sigma}_{MP}$
and ${\bar \sigma}_{ME}$ (not shown). Both of them increase  as $\tau$
is raised.
However the underdamped
case permits $\kappa$ do be adjusted  to control the dissipation rather  than the overdamped case, or even to ensure the desirable compromise between power,
efficiency and dissipation (e.g. for the parameters in Fig.~\ref{epfig}, $\tau$
and $\kappa$ can be adjusted for ensuring larger ${\cal P}$'s and lower ${\bar \sigma}$'s).  


Lastly, we address the consequences for the dissipation
at the resonant regime, as depicted by the right panels of Fig.~\ref{epfig} 
for the same parameters as in Fig.\ref{hm}. Resonance patterns are also manifested in the entropy production behavior.
Apart from the explanation in terms of Onsager coefficients, we present
an alternative  argument for the constant protocol. Since  $\eta_{MP}=1/2$ implies that
$\overline{\dot{W}}_{1} = -2\,\overline{\dot{W}}_{2}$, the entropy production acquires the simple form $\Bar{\sigma}_{MP} = \overline{\dot{W}}_{2}/T$, meaning that dissipation heat maps behaves similarly to the power heat map, albeit by a factor $T$. 
Although above similar relation is not presented for the linear protocol, resonant patterns are still visible, but less pronounced than power output and efficiency heat maps. This is because  entropy production in such cases assume values close to  those out the resonance lines.



\section{Conclusions}
\label{four}

In this paper, we introduced a sequential engine description for underdamped Brownian engines  in which the particle is subjected to distinct worksources at each 
stage. Exact thermodynamics quantities were obtained from the framework of stochastic thermodynamics. The influence of for distinct  driving driving protocols and a detailed comparison with
its overdamped dynamics was undertaken. Our findings highlighted remarkable advantages  of the underdamped dynamics when compared with its overdamped
counterpart, particularly in scenarios where minimizing dissipation while maximizing power is crucial. Despite the overdamped dynamics  exhibiting superior performances
in some specific cases, it is typically more dissipative, less efficient and presents inferior power outputs. Furthermore, the second main feature regarding underdamped case is the existence of resonance lines, which can significantly enhance the system performance. The resonance regime not only increases the power  but also enables such enhancements without compromising efficiency. This underscores the  role of such phenomena as  a mechanism of optimizing the engine operation and opens possibilities for the development of efficient and versatile Brownian engines. Overall, our comparative analysis emphasizes the importance of carefully selecting dynamics and driving protocols to optimize engine performance.

\acknowledgments 
Authors acknowledge the financial support from FAPESP under grants 2021/03372-2 and 2022/16192-5.

\appendix

\section{General Fourier coefficients of the mean velocity and General Onsager coefficients}
\label{appendix onsager}

For the underdamped case,  Fourier coefficients for the mean velocity, obtained via the solution of the system equations, are listed below:

$$a_{1kv} = \frac{8\,\pi\, k\,\tau\, \left(8\,\pi \,\gamma \,k\,\tau\, a_n+b_n\,\left(\tau ^2\, (\gamma^2 -\omega_{\textrm{D}}^{2} )-16\,\pi ^2\,k^2\right)\right)}{\left(\tau ^2\,(\gamma -\omega_{\textrm{D}})^2+16\,\pi ^2\, k^2\right) \left(\tau ^2\,(\gamma +\omega_{\textrm{D}} )^2+16\,\pi ^2\,k^2\right)},$$

$$a_{2kv} = \frac{8\,\pi\, k\,\tau\, \left(8\,\pi\, \gamma\, k\,\tau \,c_n\,+\,d_n\,\left(\tau ^2\,(\gamma^2 -\omega_{\textrm{D}}^{2} )-16\,\pi ^2\,k^2\right)\right)}{\left(\tau ^2\,(\gamma -\omega_{\textrm{D}} )^2+16\,\pi ^2 k^2\right)\,\left(\tau ^2\,(\gamma +\omega_{\textrm{D}} )^2+16\,\pi^2\,k^2\right)},$$

$$b_{1kv} = \frac{8\,\pi\, k\,\tau\, \left(a_n\,\left(\tau ^2\,\left(\omega_{\textrm{D}}^{2}-\gamma ^2\right)+16\,\pi ^2\,k^2\right)+8\,\pi\, \gamma\, k\,\tau \,b_n\right)}{\left(\tau ^2\,(\gamma -\omega_{\textrm{D}} )^2+16\,\pi ^2\,k^2\right)\,\left(\tau ^2\,(\gamma +\omega_{\textrm{D}} )^2+16\,\pi ^2\,k^2\right)},$$
and
$$b_{2kv} = \frac{8\,\pi \,k\,\tau\, \left(c_n\,\left(\tau ^2\,\left(\omega_{\textrm{D}}^i-\gamma ^2\right)+16\,\pi ^2\,k^2\right)+8\,\pi\, \gamma\, k\,\tau \,d_n\right)}{\left(\tau ^2\,(\gamma -\omega_{\textrm{D}} )^2+16\,\pi ^2\,k^2\right)\,\left(\tau ^2\,(\gamma +\omega_{\textrm{D}} )^2+16\,\pi ^2\,k^2\right)},$$
\newline
\noindent expressed in terms of Fourier coefficients  $a_n$, $b_n$, $c_n$ and $d_n$ for $g_1(t)$ and $g_2(t)$, respectively, and $\omega_{\textrm{D}} = \sqrt{\gamma^2 - 4\,\kappa}$ is the damped oscillation frequency of the system.  Onsager coefficients are related to coefficients $a_{ikv}$ and $b_{ikv}$ via the following expressions 

$$L_{11} = T\sum_{n = 1}^{\infty}\sum_{k=1}^{\infty}\,\frac{\left(\pi\, k\,a_{1kv}\,a_k-b_{1kv}\,\left(a_0\,\left((-1)^k-1\right)-\pi\, k\,b_k\right)\right)}{4\,\pi\, k} +$$ $$(1-\delta_{n,k})\cdot \left(-\frac{\left((-1)^{k+n}-1\right)\,(k\,a_n\,b_{1kv}-n\,a_{1kv}\,b_n)}{2\,\pi\, \left(k^2-n^2\right)}\right),$$

$$L_{12} = T\sum_{n = 1}^{\infty}\sum_{k=1}^{\infty}\frac{\left(\pi\, ka_{2kv}\,a_k-b_{2kv}\,\left(a_0\,\left((-1)^k-1\right)-\pi\, k\,b_k\right)\right)}{4\,\pi\, k}+$$ $$(1-\delta_{n,k})\cdot\,\left(-\frac{\left((-1)^{k+n}-1\right)\,(k\,a_n\,b_{2kv}-n\,a_{2kv}\,b_n)}{2\,\pi\, \left(k^2-n^2\right)}\right),$$

$$L_{21} = T\sum_{n = 1}^{\infty}\sum_{k=1}^{\infty}\frac{1}{4}\,\left(a_{1kv}\,c_k+\frac{c_0\,\left((-1)^k-1\right)\,b_{1kv}}{\pi\, k}+b_{1kv}\,d_k\right) + $$ $$(1-\delta_{n,k})\cdot\,\left(\frac{\left((-1)^{k+n}-1\right)\,(k\,b_{1kv}\,c_n-n\,a_{1kv}\,d_n)}{2\,\pi\, \left(k^2-n^2\right)}\right),$$
and
$$L_{22} = T\sum_{n = 1}^{\infty}\sum_{k=1}^{\infty}\frac{1}{4}\,\left(a_{2kv}\,c_k+\frac{c_0\,\left((-1)^k-1\right)\,b_{2kv}}{\pi\, k}+b_{2kv}\,d_k\right) +$$ $$(1-\delta_{n,k})\cdot\,\left(\frac{\left((-1)^{k+n}-1\right)\,(k\,b_{2kv}\,c_n-n\,a_{2kv}\,d_n)}{2\,\pi\, \left(k^2-n^2\right)}\right).$$

For constant and linear drivings, coefficients $a_n,b_n,c_n$ and $d_n$'s are listed below:
\begin{equation}
a_0=c_0=1,\qquad a_n=c_n=0, \qquad b_n=-d_n=\frac{-1+(-1)^n  }{\pi  n},
\end{equation}
for constant drivings and
\begin{equation}
a_0=c_0=\tau/4, \qquad a_n=-c_n=\frac{\left((-1)^n-1\right) \tau }{2 \pi ^2 n^2}, \qquad d_n=-\frac{ \tau }{2 \pi  n},
\end{equation}
and  $b_n=(-1)^nd_n$ for linear drivings, respectively.  For harmonic drivings, Fouriet coefficients read $a_1 = d_1 = 1/2$, $a_n = d_n= 0\,\,\forall\,\,n\neq1$, $b_1 = c_1 = 0$, $b_n = (1+(-1^n))n / ((n^2-1)\pi) = n\cdot\,c_n \,\,\forall\,\,n\neq 1$ and $c_0 = -2/\pi$.
By inserting them into expressions for $a_{1kv},a_{2kv},b_{1kv}$ and $b_{2kv}$,   Onsager coefficients can be promptly evaluated.

For the sake of completness, we also list Onsager coefficients for the overdamped case \cite{Noa1}:
\begin{equation}
L_{11}=1-\frac{2}{\gamma\tau}\tanh(\frac{\gamma\tau}{4}),
\end{equation}
and
\begin{equation}
L_{12}=\frac{2}{\gamma\tau}\tanh(\frac{\gamma\tau}{4}),
\end{equation}
for constant drivings, where $L_{11}=L_{22}$ and $L_{12}=L_{21}$ and
\begin{equation}
L_{11}=\frac{1}{12\gamma\tau}\left\{\gamma ^3 \tau ^3-3 \left(\gamma ^2 \tau ^2-8\right) \coth \left(\frac{\gamma  \tau }{2}\right)-24 \text{csch}\left(\frac{\gamma  \tau }{2}\right)\right\}
\end{equation}
and
\begin{equation}
L_{12}=\frac{\left(-\gamma  \tau +2 e^{\frac{\gamma  \tau }{2}}-2\right) \left(e^{\frac{\gamma  \tau }{2}} (\gamma  \tau -2)+2\right)}{2 \gamma  \tau  \left(e^{\gamma 
   \tau }-1\right)}
\end{equation}
for the linear drivings, respectively, where $L_{22}=L_{11}$ and $L_{12}=L_{21}$.
\section{Distinct engine operation regimes}
\label{regimeSec}
\begin{figure*}
    \centering
    \includegraphics[scale=0.5]{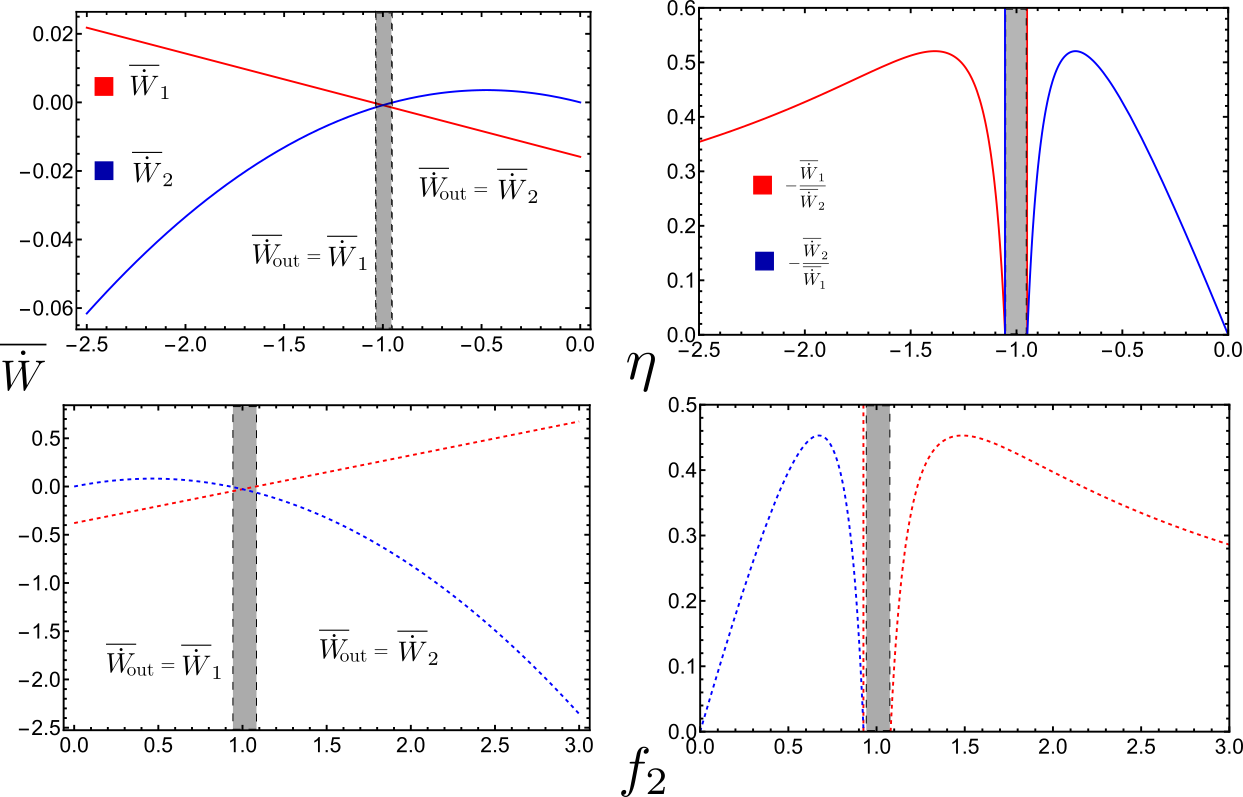}
    \caption{Depiction of  average powers $\overline{\dot{W}}_1$,$\overline{\dot{W}}_2$ (left) and efficiencies (right) for the overdamped (top continuous line) and for the underdamped (bottom dashed lines) for $f_1=1$. For better visualization, we adopted $\tau=1$ in the overdamped case and $\tau=5$ for the underdamped case. The dashed black lines represent the points where the exchange of fluxes $\overline{\dot{W}}_{\textrm{out}}\leftrightarrow\overline{\dot{W}}_{\textrm{in}} $ occurs and the gray region represents a ``dud" regime. }
    \label{trocaregime}
\end{figure*}

As discussed before in the main text, the interplay among parameters
can generate two worksources with opposite signals, $\overline{\dot{W}}_{\textrm{in}}<0$ and $\overline{\dot{W}}_{\textrm{out}}>0$, representing
the conversion of one energy (per time) into another one, the latter being identified as  the power-output.  
However, an interesting
feature in both overdamped and underdamped cases,
relies to the fact 
the increase of driving strength $f_2$ and $f_1$ held fixed 
(analogously for the other way around) 
can change the regime operation. This is depicted in Fig.~\ref{trocaregime} by plotting $\overline{\dot{W}}_{1} \leftrightarrow \overline{\dot{W}}_2$ 
as $f_2$ is changed. 
From Eqs.~(\ref{work}) and (\ref{w11}) and equal $L_{12}=L_{21}$ (the all cases here) we see that the change of regime yields at $f^*_2=\pm \sqrt{L_{11}/L_{22}}f_1$, whose efficiency is given by $\eta=-\overline{\dot{W}}_{1}/\overline{\dot{W}}_{2}$ for $f_2<f^*_2$ and $1/\eta$ for $f_2>f^*_2$, respectively. Since $L_{11}=L_{22}$ for constant and linear drivings and $f_1=1$ is held fixed, $\overline{\dot{W}}_{1}$ presents
a linear behavior, whether operating as $\overline{\dot{W}}_{\textrm{in}}$
or $\overline{\dot{W}}_{\textrm{out}}$, 
 whose crossover of regimes yields at $f^*_2=-1$ and $f^*_2=1$ for the overdamped
and underdamped cases, respectively.
As a last comment, taking into account that the efficiency is
given by the ratio between $\overline{\dot{W}}_{\textrm{out}}$ and $\overline{\dot{W}}_{\textrm{in}}$  the crossover between regimes is
characterized by infinitely large values of $\eta$ (gray lines), in similarity with previous works \cite{PhysRevResearch.5.043067}.

\section{Sinusoidal harmonic drivings}
\label{appendixper}
Sinusoidal driving forces  appears in several 
contexts, such as for  modelling Brownian particles under optical beam traps and optical tweezers \cite{PhysRevA.103.013110,ghislain1994measurement,davis2007brownian,padgett2010optical},  eletrophoresis process in colloidal gels \cite{slater1995diffusion} 
or even by measuring heat capacity experimentally
by means of oscilatting temperatures \cite{hohne2003differential,filippov1966methods}. 
In this Appendix we reproduces aformentioned features for harmonic drivings
given by
\begin{equation}
\label{sinusoidal}
    \tilde{f}_i(t) =
    \begin{cases}
       X_1 \cos(\frac{2\pi}{\tau}t)\textrm{, } 0 \leq t\leq\tau/2 \\
       X_2 \sin(\frac{2\pi}{\tau}t)\textrm{, } \tau/2 \leq t\leq \tau.
    \end{cases}
\end{equation}
%
%
\begin{figure*}[h]
    \label{senoidal_case}
    \centering
    \includegraphics[scale=0.45]{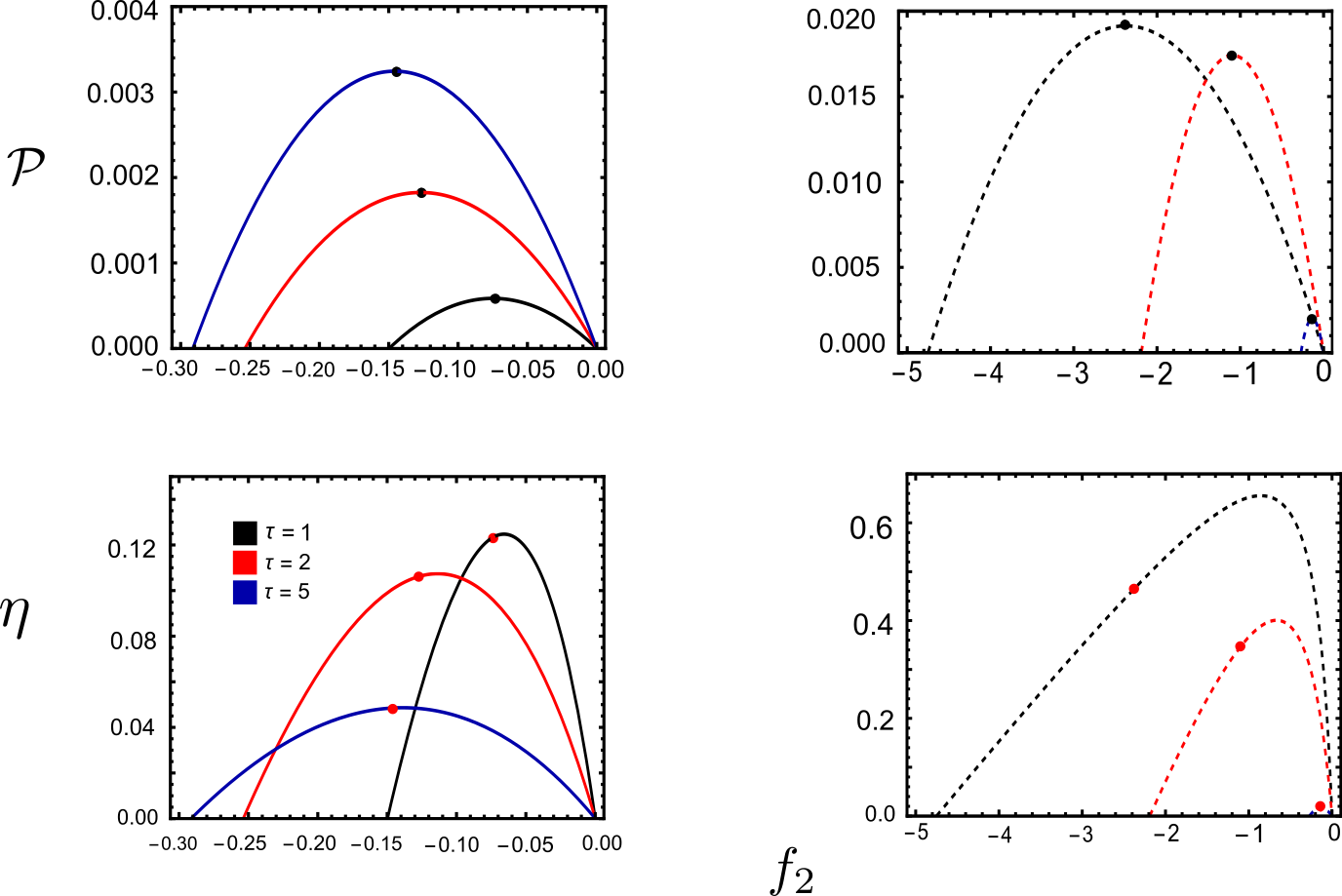}
     \caption{Power-output (top) and efficiency (bottom) versus $f_2$ for sinusoidal protocols and various periods $\tau$. Dashed lines represent the underdamped regime, while continuous lines represent the overdamped regime.  Black bullets are $\mathcal{P}_\textrm{MP}$ and red bullets are $\eta_{\textrm{MP}}$.  Parameters: $f_1 = 1$, $\gamma = 1$ and $\kappa = 1$.}
    \label{performance}
\end{figure*}

Fig.~(\ref{performance}) summarizes the main findings
for the underdamped (right) and overdamped (left) panels.
While
${\cal P}$ increases as $\tau$ is raised (overdamped), the opposite trend
is verified for the associated efficiencies and for underdamped case. In similarity with constant and linear drivings,
the underdamped case not only presents substantial larger efficiencies and power-outputs than the overdamped case but also yields a substantially larger operation regime.



\begin{figure*}
    \centering
    \includegraphics[scale=0.5]{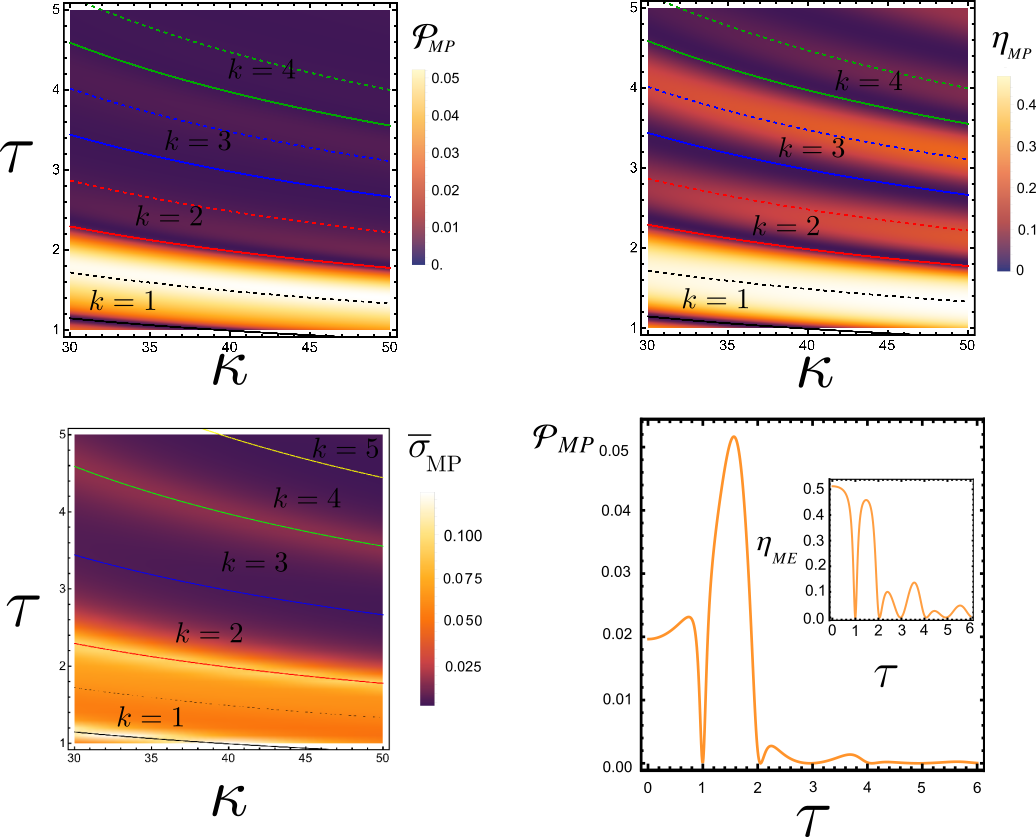}
    \caption{Depiction of ${\cal P}_{MP},\eta_{MP}$
    and ${\bar \sigma}_{MP}$ heat maps for harmonic drivings. Right bottom panels show ${\cal P}_{MP}$ and  $\eta_{ME}$ (inset)
    for $\kappa=4\pi^2$. 
    Parameters: $f_1$ and $\gamma=1=T=1$.}
    \label{hmsin}
\end{figure*}

As stated in the main text, resonant effects for  sinusoidal drivings also improves the power and efficiency, but  does not show remarkable improvements with respect to constant and linear drivings.


\bibliography{refs}

\end{document}